\documentclass[twocolumn,showpacs,amsmath,amssymb,superscriptaddress,aps,prc]{revtex4-1}
\usepackage{newlfont}
\usepackage{graphicx}
\usepackage{epstopdf}
\usepackage{epsfig}
\usepackage{bm}
\usepackage{color}

\begin{document}


\title{Experimental Determination of  $\eta$/$s$ for Finite Nuclear Matter}

\author{Debasish Mondal}
\email[e-mail:]{debasishm@vecc.gov.in}
\affiliation{Variable Energy Cyclotron Centre, 1/AF-Bidhannagar, Kolkata-700064, India}
\affiliation{Homi Bhabha National Institute, Training School Complex, Anushaktinagar, Mumbai-400094, India}

\author{Deepak Pandit}
\affiliation{Variable Energy Cyclotron Centre, 1/AF-Bidhannagar, Kolkata-700064, India}

\author{S. Mukhopadhyay}
\affiliation{Variable Energy Cyclotron Centre, 1/AF-Bidhannagar, Kolkata-700064, India}
\affiliation{Homi Bhabha National Institute, Training School Complex, Anushaktinagar, Mumbai-400094, India}

\author{Surajit Pal}
\affiliation{Variable Energy Cyclotron Centre, 1/AF-Bidhannagar, Kolkata-700064, India}

\author{Balaram Dey}
\affiliation{Tata Institute of Fundamental Research, Mumbai - 400005, India}

\author{Srijit Bhattacharya}
\affiliation{Department of Physics, Barasat Government College, Barasat, N 24 Pgs, Kolkata - 700124, India }

\author{A. De}
\affiliation{Department of Physics, Raniganj Girls' College, Raniganj-713358, India}

\author{Soumik Bhattacharya}
\affiliation{Variable Energy Cyclotron Centre, 1/AF-Bidhannagar, Kolkata-700064, India}
\affiliation{Homi Bhabha National Institute, Training School Complex, Anushaktinagar, Mumbai-400094, India}

\author{S. Bhattacharyya}
\affiliation{Variable Energy Cyclotron Centre, 1/AF-Bidhannagar, Kolkata-700064, India}
\affiliation{Homi Bhabha National Institute, Training School Complex, Anushaktinagar, Mumbai-400094, India}

\author{Pratap Roy}
\affiliation{Variable Energy Cyclotron Centre, 1/AF-Bidhannagar, Kolkata-700064, India}
\affiliation{Homi Bhabha National Institute, Training School Complex, Anushaktinagar, Mumbai-400094, India}

\author{K. Banerjee}
\affiliation{Variable Energy Cyclotron Centre, 1/AF-Bidhannagar, Kolkata-700064, India}
\affiliation{Homi Bhabha National Institute, Training School Complex, Anushaktinagar, Mumbai-400094, India}

\author{S. R. Banerjee}
\affiliation{Variable Energy Cyclotron Centre, 1/AF-Bidhannagar, Kolkata-700064, India}
\affiliation{Homi Bhabha National Institute, Training School Complex, Anushaktinagar, Mumbai-400094, India}



\begin{abstract}
We present, for the first time, simultaneous determination of shear viscosity ($\eta$) and entropy density ($s$) and thus, $\eta/s$ for equilibrated nuclear systems from $A$ $\sim$ 30 to $A$ $\sim$ 208  at different temperatures. At finite temperature, $\eta$ is estimated by utilizing the $\gamma$ decay of the isovector giant dipole resonance populated via fusion evaporation reaction, while $s$ is evaluated from the nuclear level density parameter (${a}$)  and nuclear temperature ($T$), determined precisely by the simultaneous measurements of the evaporated neutron energy spectra and the compound nuclear angular momenta. The transport parameter $\eta$ and the  thermodynamic parameter $s$ both increase with temperature resulting in a mild decrease of $\eta$/$s$ with temperature. The extracted $\eta$/$s$ is also found to be independent of the neutron-proton asymmetry at a given temperature. Interestingly, the measured $\eta$/$s$ values are comparable to that of the high-temperature quark-gluon plasma, pointing towards the fact that strong fluidity may be the universal feature of the strong interaction of many-body quantum systems.       

\end{abstract}

\maketitle

The understanding of fluidity of matter, measured by the ratio of shear viscosity ($\eta$) to entropy density ($s$), has been the subject of intense investigations in different areas of physics. The crucial ratio of $\eta/s$  is related to the Reynolds number and is well defined for both relativistic and non-relativistic fluids \cite{Scha09}. The temperature variation of $\eta/s$ also provides the crucial signature for liquid-gas phase transition in matter. String theoretical calculations have put a universal lower bound, known as the Kovtun-Son-Starinets (KSS) bound, such that $\eta/s$ $\ge$ $\hbar/4\pi k_B$ \cite{Kov05}, $k_B$ being the Boltzmann constant. In strongly coupled systems, momentum transport is highly inhibited, resulting in a small shear viscosity. The prime examples of such highly correlated systems are the Bose and the Fermi liquids \cite{Repp64, Bloc08, Gior08} at extremely low temperatures and the quark-gluon plasma (QGP), produced at high temperatures \cite{Adle03, Masu09, Aamo10}. These quantum systems have very low $\eta$/$s$ ($\sim$ 5-10 $\hbar/4\pi k_B$) \cite{Scha09} and behave as nearly perfect fluids.   

An atomic nucleus is a many-body quantum system in which the constituent particles, called nucleons, are governed by strong interaction and show highly correlated behavior. A finite nucleus, therefore, is an ideal system to search for near perfect fluidity in matter. Different model-dependent calculations for $\eta$/$s$ have been performed earlier at intermediate-energy heavy ion collisions in search for a liquid-gas phase transition \cite {Spal10, Li11, Zhou13, Fang14}.  The first theoretical study for $\eta$/$s$ in relation to the damping of giant resonances in nuclei was done by Auerbach and Shlomo \cite{Auer09} within the framework of Fermi liquid drop model (FLDM) \cite{Kolo04}. They showed that $\eta$/$s$ values for heavy and light nuclei were $\sim$ (4-19) $\hbar/4\pi k_B$ and (2.5-12.5) $\hbar/4\pi k_B$, respectively. Recently, Dang \cite{Dang11} has proposed a formalism, based on the Green-Kubo relation and the fluctuation dissipation theorem, relating the shear viscosity to the width and the energy of giant dipole resonance (GDR) in hot finite nuclei. The empirically calculated values of $\eta$/$s$ for different systems have been compared with various model calculations. A model-independent high-temperature limit of $\eta$/$s$ has also been proponed for finite nuclear systems. 

Viscosity is inherently related to the damping of the GDR, which is conceived, macroscopically, as out of phase oscillation (isovector) of the proton fluid against the neutron fluid. It is a highly damped motion characterized by a very small life time ($\sim$ 10$^{-21}$-10$^{-22}$ sec). According to the Brink-Axel hypothesis \cite{Brin55}, the GDR can be built on the ground state as well as on every excited state of the nucleus. The GDR built on the ground state  (henceforth called as the ground state GDR) is studied by photo absorption reactions, while that built on excited states is studied by fusion evaporation and inelastic scattering reactions. The line shape of the GDR is a Lorentzian, characterized by the peak energy ($E_\textrm{\scriptsize GDR}$), the width ($\Gamma_\textrm{\scriptsize GDR}$), and the resonance strength ($S_\textrm{\scriptsize GDR}$). It is observed, both experimentally and theoretically, that the $E_\textrm{\scriptsize GDR}$ and $S_\textrm{\scriptsize GDR}$ do not depend on the excitation energy ($E^*$), but $\Gamma_\textrm{\scriptsize GDR}$ increases with the increase in $E^*$. $\Gamma_\textrm{\scriptsize GDR}$ also depends on the angular momentum ($J$), owing to the $J$-induced change in shape \cite {Brac95}. However, the effect is observed above a critical angular momentum given by$J_c = 0.6A^{5/6}$ \cite{Kusn98}. The detailed discussions on giant resonances could be found in the monologues \cite{Hara01,Bort98}. It is seen that the ground state width of the GDR increases with the decrease in mass number of the nucleus. This suggests that the damping mechanism of the GDR is indeed similar to that of a viscous fluid where the modulus of decay ($\tau$) (resonance width is inversely proportional to $\tau$) of the oscillation decreases with the decrease in system volume \cite{Lamb}. In the hydrodynamic description of giant resonances proposed by Auerbach and Yeverechyahu \cite{Auer75}, the viscosity of nuclear fluids provides the mechanism for the damping of the giant states. The above description reproduces the mass dependence of the ground state GDR width, with $\eta$($T$=0) $\sim$ 1$u$, where 1$u$ = $10^{-23}$ MeV$\cdot$s$\cdot$fm$^{-3}$ \cite {Auer09,Auer75}. Microscopically, the dissipative behavior is described in terms of interparticle collisions. In case of the GDR, microscopically depicted as coherent superposition of 1$p$1$h$ states, the dissipation due to collision is formalized by taking into account the coupling of the GDR state to 2$p$2$h$ and higher order particle-hole configurations. This gives rise to the spreading width ($\Gamma^\downarrow$) of the GDR built on the ground state of the nucleus. It should be mentioned that apart from $\Gamma^\downarrow$, the ground state GDR width also comprises of Landau width ($\Gamma^\textrm{\scriptsize LD}$) and escape width ($\Gamma^\uparrow$). In case of the GDR built on excited states of atomic nuclei, the temperature-driven distortion in the nucleon momentum distribution leads to the opening of phase space, thereby creating many particle-particle ($pp$) and hole-hole ($hh$) configurations. The GDR state, apart from coupling to $ph$ configurations, also couples to these $pp$ and $hh$ configurations. This leads to the increase in the GDR width \cite {Dangn, Dangp}. An eloquent description of the increase in GDR width due to the phase space opening has also been provided in a semiclassical collision approach by Baran $et$ $al.$ \cite{Bar96}. The above discussions on the macroscopic and the microscopic approaches to the dissipation of the GDR clearly establish a qualitative  relation between the shear viscosity of the nuclear matter and the damping of the GDR owing to the two-body nucleonic interaction in nuclei.

\begin{figure}
\begin{center}
\includegraphics[height=4.9 cm, width=7.5 cm]{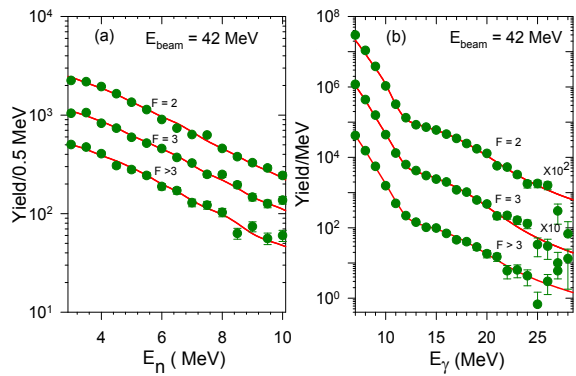}
\caption{\label{gdr} Different fold-gated experimental (symbols) (a) neutron energy spectra (b) high-energy $\gamma$-ray spectra along with the corresponding {\scshape cascade} predictions (solid lines) for $^{31}$P at $E_\textrm{\scriptsize beam}$ = 42 MeV.} 
\end{center}
\end{figure}

In this Letter, we report the first systematic simultaneous experimental determination of $\eta$, $s$, and hence, $\eta$/$s$ for equilibrated finite nuclear matter. We have utilized the prescription of Ref. \cite{Dang11} to extract $\eta$ from the measured GDR energies and widths. The entropy density $s$ has been deduced from the precisely determined nuclear temperature and temperature-dependent nuclear level density (NLD) parameter [$a(T)$], which is a measure of the single-particle density of states at the Fermi surface. The GDR was populated by light-ion-induced ($\alpha$) fusion reactions. The importance of light ion lies in the fact that the compound nuclei are populated at angular momenta ($J$), much less than the critical angular momenta for the systems. Consequently, there was no $J$-driven increase in the GDR width, and we could exclusively probe the temperature ($T$) dependence of the GDR width and thus, $\eta$.

\begin{figure}
\begin{center}
\includegraphics[height=6.0 cm, width=6.0 cm]{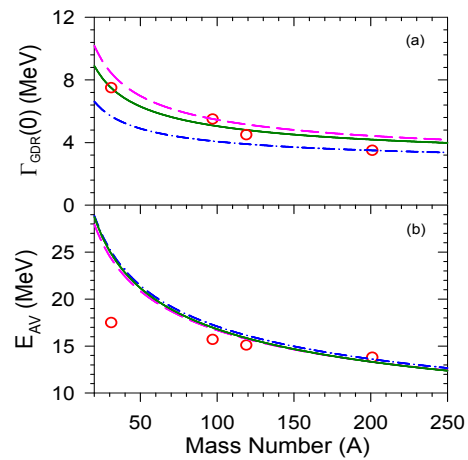}
\caption{\label{gs} Variation of the ground state (a) GDR width  and (b) average GDR energy with mass number. The symbols are (a) the accepted ground state GDR widths and (b) the average of measured GDR energies. The green solid lines are obtained by using Eqs (36), (40) and Table (3) of Ref. \cite{Auer75}, derived using $\eta(0)$ = 1$u$. The pink dashed lines and the blue dot-dashed lines were obtained by solving Eq (27) along with Eqs (25) and (28) of Ref \cite{Auer75}, with $\eta(0)$ = 1.25$u$ and 0.55$u$, respectively, by the secant method generalized for complex functions. The errors in $E_\textrm{\scriptsize AV}$ are included but are small enough to be distinguished from the symbols.} 
\end{center}
\end{figure}

The experiments were performed at the Variable Energy Cyclotron Centre (VECC), Kolkata using $\alpha$ beams from the K-130 cyclotron. The nuclei $^{31}$P, $^{97}$Tc, $^{119}$Sb and $^{201}$Tl were populated at different excitation energies in the following compound nuclear reactions: $^{4}$He ($E$$_\textrm{\scriptsize beam}$ = 28, 35, 42 MeV) + $^{27}$Al $\rightarrow$  $^{31}$P$^*$, $^{4}$He ($E$$_\textrm{\scriptsize beam}$ = 28, 35, 42, 50 MeV) + $^{93}$Nb $\rightarrow$  $^{97}$Tc$^*$, $^{4}$He ($E$$_\textrm{\scriptsize beam}$ = 30, 35, 42 MeV) + $^{115}$In $\rightarrow$  $^{119}$Sb$^*$, $^{4}$He ($E$$_\textrm{\scriptsize beam}$ = 35, 42, 50 MeV) + $^{197}$Au $\rightarrow$  $^{201}$Tl$^*$. It should be mentioned that the experimental data for $^{97}$Tc, $^{119}$Sb, and $^{201}$Tl systems have earlier been utilized to explore the low temperature variation of the GDR width \cite{Bala14,Supm12,Deep12}. The high-energy $\gamma$ rays from the decay of the GDR were measured by a part of the LAMBDA spectrometer \cite{Supm07}. Though the angular momentum populated in the compound nucleus (CN) does not affect the GDR parameters, determination of angular momentum is crucial for a precise evaluation of nuclear temperature. Hence, a 50-element multiplicity filter \cite{Deep10} was used to measure the compound nuclear angular momenta in an event-by-event mode. The cyclotron rf time spectrum was also recorded with respect to the multiplicity filter to ensure the selection of beam-related events. The angular distributions of high-energy $\gamma$ spectra were also measured for $^{31}$P and $^{119}$Sb at $E$$_\textrm{\scriptsize beam}$ = 42 MeV. Different angular-momentum-gated high-energy $\gamma$ spectra were reconstructed in the off-line analysis by the cluster summing technique \cite{Supm07}. The neutron and the pile-up events were rejected by time of flight (TOF) and pulse shape discrimination (PSD) techniques, respectively. Evaporated neutron energy spectra were also measured, in coincidence with the multiplicity $\gamma$ rays, by a liquid-scintillator-based neutron TOF detector \cite{Kaus09}. The $n-\gamma$ discrimination was accomplished following the PSD technique comprising of TOF and zero cross-over time (ZCT). The measured TOF spectra were converted to neutron energy spectra by taking the prompt $\gamma$ peak as a time reference. The details of the experiments could be found in Refs. \cite{Supm12, Bala14}.

\begin{figure}
\begin{center}
\includegraphics[height=13.0 cm, width=7.5 cm]{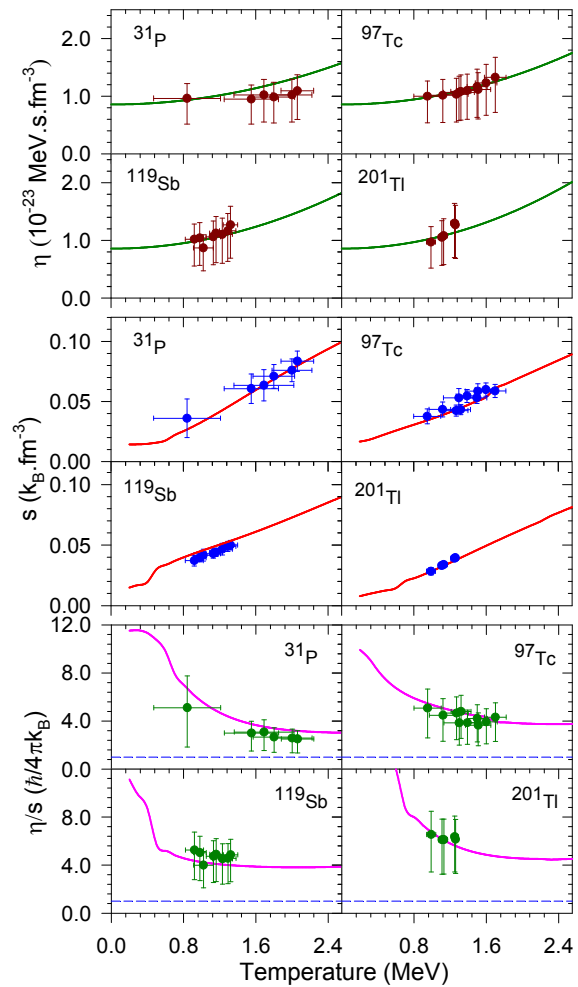}
\caption{\label{itabys} (a) Experimental data (symbols) along with the theoretical predictions (solid lines) for $\eta$ (upper panel), $s$ (middle panel), and $\eta/s$ (lower panel). Blue short-dashed line (lower panel) is the KSS bound. The errors in $\eta$ and $\eta/s$ include the statistical errors as well as the systematic error due to the lower and upper bounds of $\eta(0)$.} 
\end{center}
\end{figure}

\begin{figure}
\begin{center}
\includegraphics[height=6.9 cm, width=7.0 cm]{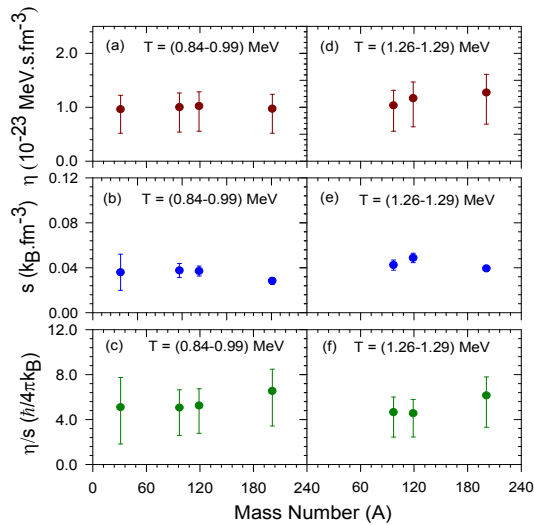}
\caption{\label{mass} Variation of $\eta$ (upper panel (a) \& (d)), $s$ (middle panel (b) \& (e)), and $\eta/s$ (lower panel (c) \& (f)) with mass number at the specified temperature range.} 
\end{center}
\end{figure}

The measured fold distributions were mapped onto the angular momentum space by a realistic technique \cite{Deep10} based on {\scshape geant3} simulations \cite {Brun86}. Different fold-gated angular momentum distributions were simulated and incorporated in a modified version of the statistical model code {\scshape cascade} \cite {Phul77}. It was shown in Ref. \cite{Prat12} that the asymptotic NLD parameter ($\tilde{a}$) depends on the angular momentum. Therefore, $\tilde{a}$  were extracted by comparing the different fold-gated neutron energy spectra, with the {\scshape cascade} predictions properly corrected for detector efficiency. Simultaneously, the calculated high-energy $\gamma$ spectra, along with a bremsstrahlung component parametrized as $\sigma = \sigma_0\exp(-E_{\gamma}/E_0)$, were folded with the detector response function and compared with the experimental spectra to extract the GDR parameters [resonance strength($S$$_\textrm{\scriptsize GDR}$), energy ($E$$_\textrm{\scriptsize GDR}$) and width ($\Gamma_\textrm{\scriptsize GDR}$)]. The center of mass (c.m.) $\gamma$-ray angular distribution was assumed to have the form $W(E_{\gamma},\theta) = W_\textrm{\scriptsize 0}(E_{\gamma})[1+a_1(E_{\gamma})P_1(\textrm{cos}\,\theta)+a_2(E_{\gamma})P_2(\textrm{cos}\,\theta)$] and the bremsstrahlung slope parameter $E$$_\textrm{\scriptsize 0}$ was deduced by comparing the experimentally measured $a_1(E_{\gamma})$ with the theoretically calculated ones. The extracted slope parameters were  consistent with the systematics $E_0 = 1.1[(E_\textrm{\scriptsize beam}-V_\textrm{\scriptsize c})/A]^{0.72}$ \cite{Nife90}. This systematics was utilized at other beam energies for which angular distributions were not measured. Different fold-gated neutron and high-energy $\gamma$-ray spectra for $^{31}$P at $E_\textrm{\scriptsize beam}$ = 42 MeV are shown in Fig. \ref{gdr} along with the {\scshape cascade} predictions.

The shear viscosity ($\eta$) was calculated at different temperatures from the measured $E_\textrm{\scriptsize GDR}$ and $\Gamma_\textrm{\scriptsize GDR}$ by utilizing the formalism of Ref. \cite{Dang11} according to which
                               
\begin{equation}
\eta(T) = \eta(0)\frac {\Gamma_\textrm{\scriptsize GDR}(T)}{\Gamma_\textrm{\scriptsize GDR}(0)}L[\Gamma_\textrm{\scriptsize GDR}(T)],
\end{equation}
where 
$L[\Gamma_\textrm{\tiny GDR}(T)] = \{\frac{E_\textrm{\tiny GDR}(0)^2}{E_\textrm{\tiny GDR}(0)^2-[\Gamma_\textrm{\tiny GDR}(0)/2]^2+[\Gamma_\textrm{\tiny GDR}(T)/2]^2}\}^2$.
%
It should be highlighted that the above prescription was deduced by incorporating a Lorentzian photo absorption cross section in the Green-Kubo relation \cite{Dang11}. We preferred the Lorentzian cross section over the Breit-Wigner cross section because the experimental GDR parameters were extracted by incorporating a Lorentzian photo-absorption cross section in the {\scshape cascade} code. As the energy of the GDR built on excited states is nearly independent of temperature, the ground state GDR energy $E_\textrm{\scriptsize GDR}(0)$ was taken as the average of measured energies ($E$$_\textrm{\scriptsize AV}$) at different temperatures while the accepted $\Gamma_\textrm{\scriptsize GDR}(0)$ for $^{31}$P, $^{97}$Tc, $^{119}$Sb and $^{201}$Tl were 7.5, 5.5, 4.5, 3.5 MeV, respectively \cite {Bala14, Deep12}. Following Refs. \cite{Auer75, Auer09, Dang11}, $\eta(0)$ was taken as 1$u$. Interestingly, the ground state GDR widths and average GDR energies (except for $^{31}$P) were well reproduced by the prescription of Ref. \cite{Auer75}, derived using $\eta(0)$ = 1$u$ (Fig. \ref{gs}). This justifies the utilization of $\eta(0)$ = 1$u$. $\eta(0)$ was also varied, according to the formalism of Ref. \cite{Auer75}, keeping all other parameters fixed as used therein, to reproduce the ground state GDR widths of $^{97}$Tc (upper bound) and $^{201}$Tl (lower bound) (see the caption of Fig. \ref{gs}). This results in the lower and the upper limits for $\eta(0)$ as 0.55$u$ and 1.25$u$, respectively. Interestingly, these bounds are quite similar to that (0.6$u$ and 1.2$u$) used in Ref. \cite{Dang11}, which were  obtained in Ref. \cite{Davi76} by comparing the calculated and the experimental most probable fission-fragment kinetic energies. These bounds have been considered as systematic errors in the deduced quantities.     

The entropy density was calculated by the relation
\begin{equation}
s(T) = \frac{\rho}{A}S(T),
\end{equation}
where the nuclear density $\rho$ = 0.16 fm$^{-3}$  and $A$ is the mass number of the nucleus. The entropy $S(T)$ is calculated following the Fermi gas model prescription as $S(T) = 2a(T)T$. $a(T)$ was deduced from the experimentally determined asymptotic NLD parameter $\tilde{a}$ by using the Ignatyuk parametrization $a(T) = \tilde{a}[1+\frac{{\Delta}S}{U}\{1-\exp(-{\gamma}U)\}]$ \cite{Igna75}. The ground state shell correction values $\Delta S$, which is the difference of experimentally measured and liquid drop masses, were -2.23, -0.20, 0.22, and -8.31 MeV for $^{31}$P, $^{97}$Tc, $^{119}$Sb, and $^{201}$Tl, respectively. These shell correction values were calculated within the {\scshape cascade} code by using the droplet model of Myers and Swiatecki \cite {Myer74} with the Wigner term. The damping factor $\gamma$, which determines the rate of shell effect depletion with excitation energy, was 0.054 MeV$^{-1}$ \cite{Prat16}. $T$ is the temperature corresponding to the excitation energy $U$ and is given by $T = \sqrt{U/a(T)}$ where $U = {E^*-E_\textrm{\scriptsize rot}-E_\textrm{\scriptsize GDR}-\Delta P}$, $E_\textrm{\scriptsize rot}$ and $\Delta P$ being the average rotational energy and the pairing energy, respectively. It should be mentioned that $\tilde{a}$ was not extracted from the neutron energy spectrum for $^{201}$Tl. The values of $\tilde{a}$ used for fitting the high-energy $\gamma$-ray spectra were utilized to extract the entropy density. Except $^{201}$Tl, $\Delta S$ is very small for all other nuclei. This results in a minute effect of $\Delta S$ and the damping factor (
$\gamma$) on deduced $a(T)$ and $T$ at the concerned nuclear excitations. However, following the recent measurements \cite{Rout13,Prat16}, an uncertainty of 0.020 MeV$^{-1}$ has been included in the damping factor. This adds to maximum systematic uncertainties of 9\% and 4\% in $a(T)$ and $T$, respectively for $^{201}$Tl at the lowest excitation. For other masses, these uncertainties are much smaller. The extracted values of $\eta$, $s$, and $\eta$/$s$ for different systems are shown in Fig. \ref{itabys} along with the respective calculations. The errors include the statistical as well as the systematic ones mentioned earlier.

Interestingly, the deduced shear viscosities are well reproduced for the systems by the calculations based on the generalized Fermi liquid drop model (FLDM) \cite{Auer09,Kolo04}. The model directly calculates $\eta$ by utilizing the two-body collisional approach and gives $\eta(T) = \frac{2}{5}\rho\epsilon_\textrm{\scriptsize F}\tau_\textrm{\scriptsize coll}/[1+(\omega\tau_\textrm{\scriptsize coll})^2]$, where $\epsilon_\textrm{\scriptsize F}$ is the Fermi energy, $\omega$ is the angular frequency of excitation, and $\tau_\textrm{\scriptsize coll}$ is the collision relaxation time given by $\tau_\textrm{\scriptsize coll} = \tau_0/[1+(\hbar\omega/2\pi T)^2]$, $\tau_0 = \hbar\alpha/T^2$. In the rare collision (zero sound) regime (which corresponds to giant resonances) $\omega\tau$ $\gg$ 1 and at low temperatures such that $T \ll \hbar\omega$, $\eta$ reduces to $\eta = \frac{2}{5}\rho\epsilon_F\frac{\hbar}{4\pi^2\alpha}\times[1+(2\pi T/\hbar\omega)^2]$. The parameter $\alpha$ depends on the in-medium nucleon-nucleon scattering cross section and for isovector resonances its value is 4.6 MeV \cite{Kolo04}. The theoretical results are obtained using the values of $\epsilon_\textrm{\scriptsize F}$ = 37 MeV, corresponding to $\rho$ = 0.16 fm$^{-3}$ and considering $\hbar\omega$ as the average GDR energy. It is observed that at low temperatures, $\eta$ increases with $T$, which can be understood qualitatively by the following arguments. For an equilibrated nucleus, the momentum is transported by the nucleons. The kinetic theoretical calculations give $\eta = \frac{1}{3}\rho m \bar{v}\lambda$, where $\bar{v}$ is the average velocity of the nucleons, and the mean free path $\lambda \sim \bar{v}/N_\textrm{\scriptsize coll}$. In the rare collision region, the collision frequency $N_\textrm{\scriptsize coll}$ does not change much with temperature, while $\bar{v} \sim \sqrt{T}$. Therefore, the mean free path as well as the nucleon momentum increase with temperature. That means the momentum can be transported more efficiently over a large distance, thereby increasing $\eta$ with temperature. It has been observed experimentally in all mass regions that the GDR width remains constant up to some critical temperature ($T_c$ = 0.7+37.5/$A$) and increases thereafter \cite {Deep12,Bala14}. On the other hand, FLDM predicts a gradual increase of the GDR width from $T$ = 0 \cite{Dang11}. This could be a possible reason for the discrepancy in $\eta(T)$ for $^{31}$P as most of the data are below or near $T_c$ (1.9 MeV). However, for other masses, most of the data points are above their respective $T_c$, resulting in good agreement between theory and experiment. It is also appealing to note that the measured entropy density is well reproduced by the calculations. $s(T)$ is estimated utilizing the relation $s(T) =  \frac{\rho}{A}\times[-\sum_i f_i\ln(f_i)-\sum_i(1-f_i)\cdot\ln(1-f_i)]$, where $f_i$ is the Fermi function given by $f_i = [1+\exp\{(e_i-\mu)/T\}]^{-1}$. The chemical potential $\mu$ is calculated from particle conservation, and the single particle energies $e_i$ are calculated using the deformed Wood-Saxon potential with the universal parameters \cite{Cwio87}. As the temperature increases, the distortion of the Fermi surface becomes larger, thereby increasing the number of accessible microstates. This results in the increase of entropy density with temperature.

The deduced $\eta$/$s$ shows (Fig. \ref{itabys} ) a mild decrease with temperature. Moreover, it is confined in the range (2.5-6.5) $\hbar/4\pi k_B$ for the finite nuclear matter within the temperature range $\sim$ (0.8-2.1) MeV. Interestingly, the measured $\eta$/$s$ is comparable to that of the QGP \cite{Scha09}. It, therefore, could be reaffirmed, as pointed out in Ref. \cite {Auer09}, that the strong fluidity is a universal characteristic feature of the strong interaction of the many-body nuclear systems and not just of the state created in the relativistic collisions. It is also fascinating to note that, although $\eta$ shows a slight increase with the mass number (Fig. (\ref{mass}d)) at the highest available temperature for heavier nuclei, $\eta$/$s$ remains within (5.1-6.5) $\hbar/4\pi k_B$ (Fig. (\ref{mass}c)) and (4.6-6.1) $\hbar/4\pi k_B$ (Fig. (\ref{mass}f)) at the lowest and highest available temperatures, respectively, for all nuclei. This indicates that $\eta$/$s$ is approximately independent of the nuclear size and the neutron-proton asymmetry at a given temperature. However, it could be the artefact of incorporating the same $\eta(0)$ for all nuclei. Also owing to large errors, the data are not sensitive enough to draw any conclusion and thus, call for further studies. It will also be interesting to extend this study to the limiting temperature of existence of the GDR to further investigate the high-temperature limit of $\eta$/$s$ as proposed in Ref. \cite{Dang11}.

In summary, we have experimentally determined, for the first time, the ratio of shear viscosity to entropy density for the finite nuclear matter at different temperatures. $\eta$ was extracted from the measured isovector giant dipole resonance energy and width, while $s$ was deduced from the precisely determined nuclear temperature and nuclear level density parameter. Both $\eta$ and $s$ increase with temperature, resulting in a mild decrease in $\eta$/$s$ with temperature. At a given temperature, $\eta$/$s$ is also found to be approximately independent of the nuclear size as well as the neutron-proton asymmetry. Moreover, the measured $\eta$/$s$ remains confined in the range (2.5-6.5) $\hbar/4\pi k_B$ and thus, establishes that strong fluidity is the universal characteristic of the strong interaction of the many-body quantum systems.                                                     

The authors sincerely acknowledge the stimulating discussions with Professor Dinesh K. Srivastava and Professor Jan-e Alam. The authors also appreciate the discussions with Dr. Shashi C. L. Srivastava. The authors are thankful to VECC cyclotron operators for smooth running of the accelerator during the experiment.

\end{document}